# Accessing the Full Superconducting Dome in Pristine YBa$_2$Cu$_3$O$_{6+x}$ Under Pressure

P. L. Alireza,[1] G. H. Zhang,[1,2] W. Guo,[1] J. Porras,[3] T. Loew,[3] Y. -T. Hsu,[1]
G. G. Lonzarich,[1] M. Le Tacon,[3,4] B. Keimer,[3] and Suchitra E. Sebastian[1, *]

[1]*Cavendish Laboratory, Cambridge University, JJ Thomson Avenue, Cambridge CB3 OHE, U.K*
[2]*Department of Physics, Massachusetts Institute of Technology, Cambridge, Massachusetts 02139, USA*
[3]*Max-Planck-Institut für Festkörperforschung, Heisenbergstrasse 1, D-70569 Stuttgart, Germany*
[4]*Karlsruher Institut für Technologie, Institut für Festkörperphysik,
Hermann-v.-Helmholtz-Platz 1, D-76344 Eggenstein-Leopoldshafen, Germany*
(Dated: October 29, 2016)

We attain the previously unaccessed full superconducting dome in a pristine high temperature cuprate superconductor by applying pressures up to 280 kbar to samples of near stoichiometric YBa$_2$Cu$_3$O$_7$. The obtained superconducting phase boundary as a function of hole doping closely follows the superconducting dome in La$_{2-x}$Sr$_x$CuO$_4$. Measurements are now enabled to trace the evolution of various entangled phases and the Fermi surface from the underdoped to overdoped regime in a single high purity cuprate superconducting family of materials.

Copper oxide high temperature superconductors, while showing apparently conventional Fermi liquid behaviour in the overdoped region of the phase diagram, exhibit a mysterious plethora of phases in the underdoped region of the phase diagram. In addition to $d$-wave superconductivity and Mott insulating antiferromagnetism, phases identified in the underdoped region include various forms of charge density wave, spin density wave, electronic nematic order, and a puzzling pseudogap region [1]. Recent studies have also suggested the potential deviation of superconducting behaviour in the overdoped regime from the conventional Bardeen-Cooper-Schrieffer (BCS) theory [2, 3]. It is as yet an open question as to the relevance of each of the identified superconducting and density wave phases to the pseudogap region, and the evolution of each of these phases from the underdoped to the overdoped Fermi-liquid-like region. An understanding of this evolution will shed light on the nature of the ill-understood pseudogap region, yet this has proved challenging thus far given the unavailability of pristine cuprate materials families that can be tuned across the entire superconducting phase diagram from underdoped to overdoped.

Among the high-$T_\mathrm{c}$ superconductors, YBa$_2$Cu$_3$O$_{6+x}$ (YBCO) (see inset of figure 1), is a material recognized for being clean and well-ordered, which leads to its prevalence in quantum oscillation experiments and other measurements [4, 5]. It has become one of the most extensively studied materials in the cuprate family, having well-identified critical points [6] and electronic orders [7–11]. While quantum oscillations and transport have been thoroughly studied in underdoped YBCO [12–14], similar measurements cannot be extended to the overdoped regime in the same material, despite having been successfully used to study the Fermi surface of a different cuprate, Tl$_2$Ba$_2$CuO$_{6+x}$ (Tl-2201) [15, 16]. This is due to the fact that chemical doping in YBCO can only achieve a maximum hole concentration of 19.4% with oxygen tuning, and up to 22% with calcium substitution [17]. It is important to find a control parameter that can tune YBCO from the underdoped to the overdoped region in a pristine way in order to track the evolution of the various forms of order and their relation to the pseudogap across the full range of doping, as well as to investigate the overdoped region itself. Hydrostatic pressure has been used as a tuning parameter [18, 19] in past experiments to tune between adjoining electronic phases in materials where the doping or bandwidth was found to change significantly with pressure [20]. For instance, pressure has both induced superconductivity and suppressed competing orders in magnetic materials such as UGe$_2$ [21–26], and revealed in materials such as CePd$_2$Si$_2$ and CeIn$_3$ critical points as a function of pressure (doping) that are relevant to superconductivity [27, 28]. We report that by starting near YBa$_2$Cu$_3$O$_7$ and applying high pressures of up to 280 kbar, we have tuned the material into the heavily overdoped region up to a hole doping of $\approx 26\%$, where the superconducting transition temperature becomes almost fully suppressed. By using applied pressure as the control parameter in YBCO, therefore, we have achieved unprecedented access to the entire superconducting dome in a pristine high temperature superconducting material.

Previous studies of YBCO under lower applied pressures have shown that increasing hydrostatic pressure decreases the distance between the BaO and CuO$_2$ planes, compressing the softer Cu(plane)-O bond and elongating the stiffer Cu(chain)-O bond (see inset of figure 1), and thus increasing the number of holes per Cu atom in the CuO$_2$ plane ($n_\mathrm{h}$) through hole transfer from the CuO chains [29, 30]. While pressure has also been suggested to have intrinsic effects on $T_\mathrm{c}$ through induced structural phase transitions, oxygen ordering and changes in the effective interaction strength [31–33], such effects are in practice expected to be negligible above the slightly overdoped region [34–36]. Thus, we assume that in the region which is of interest to us, from the optimally doped region to the overdoped region, the effect of pressure is principally to tune the properties of YBCO through dop-ing. Recent experiments have shown pressure to be an effective control parameter for cuprates; resistivity measurements in $Bi_{1.98}Sr_{2.06}Y_{0.68}CaCu_2O_{8+x}$ have demonstrated pressure-induced superconductivity [37], while pressure-tuned Raman and x-ray diffraction data have indicated quantum critical points corresponding to electronic transitions [38]. For YBCO specifically, hydrostatic pressure has been applied up to 170 kbar on an initially underdoped sample with $x = 0.66$ and up to 110 kbar on an initially optimally doped sample with $x = 0.95$, showing pressure to affect the superconducting critical temperature in both cases. In the former, the effects of pressure, applied in the underdoped region, are complicated by oxygen ordering and thus difficult to analyze while the latter achieves a maximum effective $n_h$ of around 0.2 which ventures into the slightly overdoped region [34, 35]. Here, we start near stoichiometric $YBa_2Cu_3O_7$ in which the hole fraction is intrinsically just above optimal, in order to most effectively access the highly overdoped regime with applied pressure as a tuning parameter.

Single crystals of $YBa_2Cu_3O_{6+x}$ with $x = 0.98$ were grown out of solution [39] and detwinned under uniaxial pressure. The nominal oxygen content was determined by relating the superconducting temperature to the hole doping and oxygen concentration using the previously determined relationship for this material [17]. Polycrystalline $YBa_2Cu_3O_7$ was grown by conventional solid state reaction followed by sintering and cooling, according to methods described in previous literature [40]. To determine the effect of pressure on the superconducting transition temperature, we measured ac-susceptibility in a diamond anvil cell (DAC) using a micro-coil system [41]. Pressures of up to 300 kbar were generated using 0.6 mm culet diamonds with glycerol as the pressure medium. The sample space was a 200 $\mu$m hole drilled into a MP35 gasket, in which the signal coil, a 130 $\mu$m diameter, 4 turn micro-coil made with 10 $\mu$m insulated copper wire, was placed. The drive coil, a 130 turn coil made with 30 $\mu$m insulated copper wire, was placed outside the pressure region. Pressure inside the sample space was determined by the ruby fluorescence method at room temperature and in some instances verified by the superconducting transition temperature of a lead sample at low temperatures. Additional measurements on DC magnetization were performed using a SQUID anvil cell specifically developed to be used in a Quantum Design MPMS system [42, 43]. The cell, gasket, and pressure medium were set up similarly to the one described above, using 0.6 mm diamond anvils, MP35 gasket, and glycerol as the pressure medium.

Ac-susceptibility curves measured for both single crystal and polycrystalline YBCO at several values of applied pressure are shown in figure 1, where the susceptibilities have been normalised. We define the superconducting temperature as corresponding to the midpoint of the ac-susceptibility curve. We find, as shown in table I and figure 2, that up to applied pressures of around 90 kbar, the evolution in superconducting temperature is modest, falling by $\approx 20\%$, similar to that previously reported [35]. However, at higher values of applied pressures beyond those previously reported, the superconducting transition temperature begins to drop rapidly. For significantly high pressure up to 280 kbar, we find that superconductivity is almost destroyed, with the superconducting temperature falling to below 10 K. It is clear that in the case of YBCO, the large energy scales that characterise this material necessitate extremely high applied pressures far beyond those achieved in previous studies to access the overdoped region of the phase diagram. Superconducting transition temperatures as a function of applied pressure are found to be similar for single crystal and polycrystalline YBCO, demonstrating the suppression of superconductivity to be primarily a volume effect rather than a uniaxial effect. The width of the superconducting transition under pressure reflects both the gasket condition during the run and the level of pressure homogeneity within the sample. The remarkable sharpness of the superconducting transitions we observe at the highest applied pressures in the vicinity of 250 kbar (figure 1) demonstrates the high homogeneity of pressure conditions achieved even at these very high pressures.

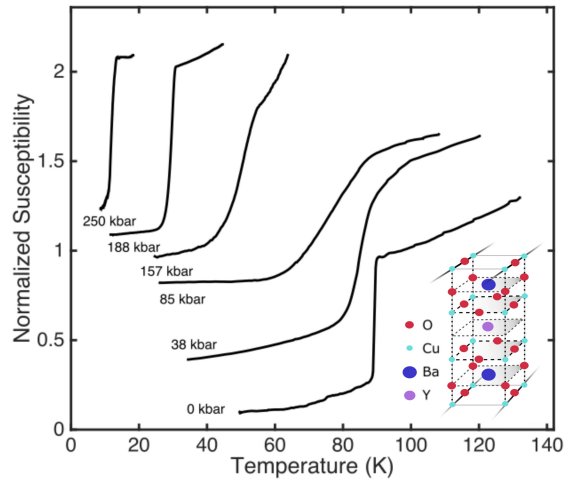

FIG. 1. Progression of ac-susceptibility curves for single crystal $YBa_2Cu_3O_{6.98}$ with increasing applied pressure. Measurements were taken over multiple runs, by increasing pressure in the same setup or by remounting the sample at a higher pressure in a new setup. The varied widths of each transition are likely due to specific conditions of the setup at each run, including pressure homogeneity and potential deformation of the gasket. The inset shows the YBCO unit cell.

| Single Crystal YBCO | | Polycrystalline YBCO | |
|---|---|---|---|
| Pressure (kbar) | $T_c$ (K) | Pressure (kbar) | $T_c$ (K) |
| 0 | 89.1 ±0.2 | 5 | 84.8 ±0.3 |
| 38 | 85.1 ±0.9 | 17 | 85.5 ±0.3 |
| 85 | 76 ±2 | 34 | 83.7 ±0.6 |
| 90 | 70 ±2 | 64 | 83.7 ±0.9 |
| 113 | 68 ±3 | 90 | 72 ±2 |
| 157 | 48.7 ±0.2 | 143 | 66 ±5 |
| 188 | 29 ±2 | 172 | 49 ±6 |
| 204 | 28.0 ±0.2 | 25 | 43 ±2 |
| 250 | 12 ±1 | 240 | 23.4 ±0.6 |
| 280 | 7.34 ±0.04 | 270 | 8.7 ±0.2 |

TABLE I. Superconducting temperatures obtained from ac-susceptibility measurements on YBCO under several values of applied pressure, obtained from figure 1. The left side of the table shows results on single crystal YBCO and the right table shows results on polycrystalline YBCO, which are found to be similar.

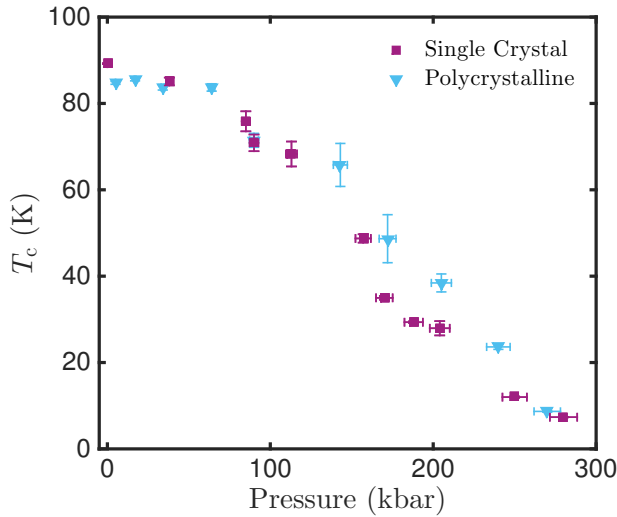

FIG. 2. Determination of $T_c$ from ac-susceptibility curves shows the decrease of transition temperature with increasing pressure for both single crystal and polycrystalline YBCO. Each point has a horizontal error of $\pm 2$ kbar or $\pm 3\%$, the greater of which indicates uncertainty in the determination of pressure inside the sample space as described above. The vertical error bar corresponds to uncertainty in $T_c$, which is related to the width of the transition.

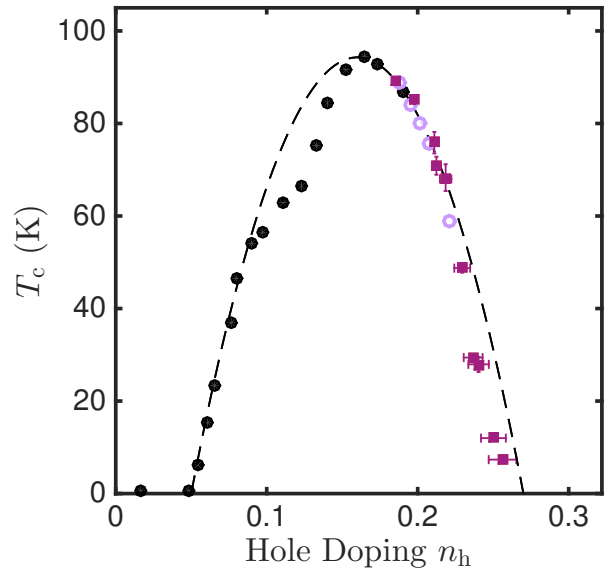

FIG. 3. $T_c$-doping phase diagram for single crystal YBCO, with translated pressure-tuned data from figure 2 (square), previous Calcium-doped data (hollow circle) [44], and previous zero-pressure data (filled circle) [17]. Translation from pressure to doping $n_h$ was done using relations 1 and 2 while considering constraints in accordance with previous data on $Y_{1-x}Ca_xBa_2Cu_3O_{6.96}$ [44], for which $n_h$ was estimated using a combination of copper and oxygen bond valence sums (BVS). The effects of the different constraints are reflected in the horizontal error bars, and the dashed curve indicates the LSCO superconducting dome, $1 - T_c/T_{c,\max} = 82.6(n_h - 0.16)^2$, with $T_{c,\max} = 94.3$ K for YBCO single crystals [17].

In order to map the YBCO phase diagram with $T_c$ as a function of $n_h$ for our applied pressure measurements, we need to translate applied pressure into hole doping. Early pressure studies have shown the effect of pressure on the change in $n_h$ to closely correspond to a linear dependence on the relative decrease in cell volume [45–47]. We follow similar analysis; denoting the relative decrease in the unit cell volume as $\xi(P) \equiv (1 - V(P)/V_0)$, where $V$ is the cell volume, $P$ is the applied pressure, and $V_0$ is the cell volume at zero pressure. Using the first-order Murnaghan equation [48], we assume,

$$\frac{dn_h(P)}{d\xi(P)} = n'_h(P)B_0\left(1 + \frac{B'_0}{B_0}P\right)^{1+1/B'_0} \quad (1)$$

$$\xi(P) = 1 - \left(1 + \frac{B'_0}{B_0}P\right)^{-1/B'_0}, \quad (2)$$

where $n'_h(P) = dn_h(P)/dP$ and $n_{h,0}$, $B_0$, and $B'_0$ are the hole doping, bulk modulus $-V(dP/dV)$, and pressure derivative of the bulk modulus respectively at zero pressure. The initial doping $n_h(0) = 0.19$ holes/Cu was found by locating $T_c$ on the LSCO superconducting parabola, $1 - T_c/T_{c,\max} = 82.6(n_h - 0.16)^2$ with $T_{c,\max} = 94.3$ K for YBCO single crystals [17]. Values considered for $B_0$ and $B'_0$ were taken from previous pressure studies using X-ray analysis, on samples with oxygen content close to that of our single crystals $x = 0.98$; as such, we assume a range of $B_0 = 1306 \pm 11$ kbar and $B'_0 = 4.8 \pm 0.2$ to $B_0 = 1500$ kbar and $B'_0 = 4$, obtained for $x = 0.95$ [49] and $x = 1$ [50], respectively. Starting with an assumption of parabolic dependence of the superconducting temperature as a function of hole doping up to $n_h = 0.22$ similar to previous work [44], we apply relations 1 and 2 (See Supplementary Material for further details of the calculation) to yield the dependence of the superconducting temperature as a function of hole doping at higher dopings. The full superconducting phase diagram thus obtained as a function of hole doping is shown in figure 3. We are therefore able to map the full superconducting dome in a pristine way in a high temperature cuprate superconductor by applied pressure tuning starting near stoichiometric $YBa_2Cu_3O_7$. Importantly, we find that the superconducting temperature dependence on hole doping is in good agreement with the form of the superconducting dome mapped in the LSCO cuprates [51–53].

Crucially, our applied pressure measurements provide unprecendented access beyond the optimally doped region in YBCO, allowing for measurements of quantum oscillations, electrical and thermal transport, magnetisation and other thermodynamic properties in the overdoped regime of this family of high-$T_c$ materials. In particular, we expect to trace the evolution of the Fermi surface pockets from small to large across the entire phase diagram, and to resolve each of the entangled density wave and $d$-wave superconducting phases from the underdoped to the overdoped region, and their relation to each other and the pseudogap region in this family of high-$T_c$ cuprates. Another important question to be addressed pertains to the robustness of superconductivity with increasing temperature and applied magnetic field in the underdoped region compared to the overdoped region. Controversy surrounds whether superconductivity persists to high magnetic fields in the underdoped regime, as would be characteristic of strongly coupled superconductivity in proximity to the Mott transition [54], or whether superconductivity in the underdoped regime is suppressed by modest critical magnetic fields as would be characteristic of conventional BCS superconductivity [55]. We can shed light on this issue by a comparative study of the robustness of superconductivity in the overdoped regime of YBCO we now access, which is expected to be closer in behaviour to conventional BCS superconductivity.

P.L.A., G.H.Z., W.G., Y.T.H., and S.E.S. acknowledge support from the Royal Society, the Winton Programme for the Physics of Sustainability, and the European Research Council under the European Union's Seventh Framework Programme (grant number FP/2007-2013)/ERC Grant Agreement number 337425. G.H.Z. acknowledges support from the Cambridge-MIT Exchange Program. G.G.L. acknowledges support from EPSRC grant EP/K012894/1.

---

* ses59@cam.ac.uk
[1] B. Keimer, S. A. Kivelson, M. R. Norman, S. Uchida, and J. Zaanen, Nature **518**, 179 (2015).
[2] I. Božović, X. He, J. Wu, and A. T. Bollinger, Nature **536**, 309 (2016).
[3] D. C. Peets, D. G. Hawthorn, K. M. Shen, Y.-J. Kim, D. S. Ellis, H. Zhang, S. Komiya, Y. Ando, G. A. Sawatzky, R. Liang, D. A. Bonn, and W. N. Hardy, Physical Review Letters **103**, 087402 (2009).
[4] S. E. Sebastian and C. Proust, *Annual Review of Condensed Matter Physics*, Annual Review of Condensed Matter Physics **6**, 411 (2015).
[5] S. E. Sebastian, N. Harrison, and G. G. Lonzarich, Reports on Progress in Physics **75**, 102501 (2012).
[6] J. L. Tallon and J. W. Loram, Physica C: Superconductivity **349**, 53 (2001).
[7] D. Fournier, G. Levy, Y. Pennec, J. L. McChesney, A. Bostwick, E. Rotenberg, R. Liang, W. N. Hardy, D. A. Bonn, I. S. Elfimov, and A. Damascelli, Nat Phys **6**, 905 (2010).
[8] T. Wu, H. Mayaffre, S. Kramer, M. Horvatic, C. Berthier, W. N. Hardy, R. Liang, D. A. Bonn, and M.-H. Julien, Nature **477**, 191 (2011).
[9] M. Hücker, N. B. Christensen, A. T. Holmes, E. Blackburn, E. M. Forgan, R. Liang, D. A. Bonn, W. N. Hardy, O. Gutowski, M. v. Zimmermann, S. M. Hayden, and J. Chang, Physical Review B **90**, 054514 (2014).
[10] S. Blanco-Canosa, A. Frano, E. Schierle, J. Porras, T. Loew, M. Minola, M. Bluschke, E. Weschke, B. Keimer, and M. Le Tacon, Physical Review B **90**, 054513 (2014).
[11] T. Wu, H. Mayaffre, S. Krämer, M. Horvatić, C. Berthier, P. L. Kuhns, A. P. Reyes, R. Liang, W. N. Hardy, D. A. Bonn, and M.-H. Julien, Nat Commun **4** (2013).
[12] S. E. Sebastian, N. Harrison, F. F. Balakirev, M. M. Altarawneh, P. A. Goddard, R. Liang, D. A. Bonn, W. N. Hardy, and G. G. Lonzarich, Nature **511**, 61 (2014).
[13] S. Badoux, W. Tabis, F. Laliberté, G. Grissonnanche, B. Vignolle, D. Vignolles, J. Béard, D. A. Bonn, W. N. Hardy, R. Liang, N. Doiron-Leyraud, L. Taillefer, and C. Proust, Nature **531**, 210 (2016).
[14] Y. Ando, Y. Kurita, S. Komiya, S. Ono, and K. Segawa, Phys. Rev. Lett. **92**, 197001 (2004).
[15] B. Vignolle, A. Carrington, R. A. Cooper, M. M. J. French, A. P. Mackenzie, C. Jaudet, D. Vignolles, C. Proust, and N. E. Hussey, Nature **455**, 952 (2008).
[16] A. F. Bangura, P. M. C. Rourke, T. M. Benseman, M. Matusiak, J. R. Cooper, N. E. Hussey, and A. Carrington, Physical Review B **82**, 140501 (2010).
[17] R. Liang, D. A. Bonn, and W. N. Hardy, Phys. Rev. B **73**, 180505 (2006).
[18] G. G. Lonzarich, in *Electron*, edited by M. Springford (Cambridge University Press, 1997) Chap. 6.
[19] H. Fukuyama, The Review of High Pressure Science and Technology **7**, 465 (1998).
[20] F. M. Grosche, S. R. Julian, N. D. Mathur, and G. G. Lonzarich, *Proceedings of the International Conference on Strongly Correlated Electron Systems*, Physica B: Condensed Matter **223–224**, 50 (1996).
[21] N. D. Mathur, F. M. Grosche, S. R. Julian, I. R. Walker, D. M. Freye, R. K. W. Haselwimmer, and G. G. Lonzarich, Nature **394**, 39 (1998).
[22] S. S. Saxena, P. Agarwal, K. Ahilan, F. M. Grosche, R. K. W. Haselwimmer, M. J. Steiner, E. Pugh, I. R. Walker, S. R. Julian, P. Monthoux, G. G. Lonzarich, A. Huxley, I. Sheikin, D. Braithwaite, and J. Flouquet, Nature **406**, 587 (2000).
[23] G. Oomi, T. Kagayama, and Y. Onuki, J. Alloys Compounds **271-273**, 482 (1998).

[24] K. Nishimura, G. Oomi, S. W. Yun, and Y. Onuki, J. Alloys Compounds **213**, 383 (1994).
[25] A. Huxley, I. Sheikin, and D. Braithwaite, Physica B **284 & 288**, 1277 (2000).
[26] M. Leroux, I. Errea, M. Le Tacon, S.-M. Souliou, G. Garbarino, L. Cario, A. Bosak, F. Mauri, M. Calandra, and P. Rodière, Physical Review B **92**, 140303 (2015).
[27] J. D. Thompson, R. D. Parks, and H. Borges, J. Magn. Magn. Mater. **54-57**, 377 (1986).
[28] J. Flouquet, J. Magn. Magn. Mater. **90 & 91**, 377 (1990).
[29] J. Jorgensen, S. Pei, P. Lightfoot, D. Hinks, B. Veal, B. Dabrowski, A. Paulikas, R. Kleb, and I. Brown, Physica C: Superconductivity **171**, 93 (1990).
[30] Y. Yamada, J. D. Jorgensen, S. Pei, P. Lightfoot, Y. Kodama, T. Matsumoto, and F. Izumi, Physica C: Superconductivity **173**, 185 (1991).
[31] S. W. Tozer, J. L. Koston, and E. M. McCarron III, Physical Review B **47**, 8089 (1993).
[32] J. Schilling, in *Handbook of High Temperature Superconductivity: Theory and Experiment*, edited by J. Schrieffer and J. Brooks (Springer Verlag, Hamburg, 2007) Chap. 11.
[33] S. M. Souliou, A. Subedi, Y. T. Song, C. T. Lin, K. Syassen, B. Keimer, and M. Le Tacon, Physical Review B **90**, 140501 (2014).
[34] O. Cyr-Choinière, D. LeBoeuf, S. Badoux, S. Dufour-Beauséjour, D. A. Bonn, W. N. Hardy, R. Liang, N. Doiron-Leyraud, and L. Taillefer, ArXiv e-prints (2015), arXiv:1503.02033 [cond-mat.supr-con].
[35] S. Sadewasser, J. S. Schilling, A. P. Paulikas, and B. W. Veal, Physical Review B **61**, 741 (2000).
[36] B. Lorenz and C. Chu, in *Frontiers in Superconducting Materials*, edited by A. Narlikar (Springer Verlag, Berlin, 2005) pp. 459–497.
[37] T. Cuk, D. A. Zocco, H. Eisaki, V. Struzhkin, F. M. Grosche, M. B. Maple, and Z. X. Shen, Physical Review B **81**, 184509 (2010).
[38] T. Cuk, V. V. Struzhkin, T. P. Devereaux, A. F. Goncharov, C. A. Kendziora, H. Eisaki, H.-K. Mao, and Z.-X. Shen, Phys. Rev. Lett. **100**, 217003 (2008).
[39] C. T. Lin, W. Zhou, W. Y. Liang, E. Schönherr, and H. Bender, Physica C: Superconductivity **195**, 291 (1992).
[40] S. Zagoulaev, P. Monod, and J. Jégoudez, Phys. Rev. B **52**, 10474 (1995).
[41] P. L. Alireza and S. R. Julian, Review of Scientific Instruments **74**, 4728 (2003).
[42] P. L. Alireza and G. G. Lonzarich, Review of Scientific Instruments **80**, 023906 (2009), http://dx.doi.org/10.1063/1.3077303.
[43] P. L. Alireza, S. Barakat, A.-M. Cumberlidge, G. Lonzarich, F. Nakamura, and Y. Maeno, Journal of the Physical Society of Japan **76**, 216 (2007), http://dx.doi.org/10.1143/JPSJS.76SA.216.
[44] J. L. Tallon, C. Bernhard, H. Shaked, R. L. Hitterman, and J. D. Jorgensen, Phys. Rev. B **51**, 12911 (1995).
[45] X.-J. Chen, V. V. Struzhkin, R. J. Hemley, H.-k. Mao, and C. Kendziora, Phys. Rev. B **70**, 214502 (2004).
[46] X. J. Chen, H. Q. Lin, and C. D. Gong, Physical Review Letters **85**, 2180 (2000).
[47] W. H. Fietz, F. W. Hornung, K. Grube, S. I. Schlachter, T. Wolf, B. Obst, and P. Schweiss, Journal of Low Temperature Physics **117**, 915 (1999).
[48] J. S. Olsen, S. Steenstrup, L. Gerward, and B. Sundqvist, Physica Scripta **44**, 211 (1991).
[49] I. V. Medvedeva, Y. S. Bersenev, B. A. Gizhevsky, N. M. Chebotaev, S. V. Naumov, and G. B. Demishev, Zeitschrift für Physik B Condensed Matter **81**, 311.
[50] H. Ludwig, W. Fietz, and H. Wühl, Physica C: Superconductivity **197**, 113 (1992).
[51] J. B. Torrance, A. Bezinge, A. I. Nazzal, T. C. Huang, S. S. P. Parkin, D. T. Keane, S. J. LaPlaca, P. M. Horn, and G. A. Held, Physical Review B **40**, 8872 (1989).
[52] H. Takagi, T. Ido, S. Ishibashi, M. Uota, S. Uchida, and Y. Tokura, Physical Review B **40**, 2254 (1989).
[53] M. Presland, J. Tallon, R. Buckley, R. Liu, and N. Flower, Physica C: Superconductivity **176**, 95 (1991).
[54] L. Li, Y. Wang, S. Komiya, S. Ono, Y. Ando, G. D. Gu, and N. P. Ong, Physical Review B **81**, 054510 (2010).
[55] G. Grissonnanche, O. Cyr-Choinière, F. Laliberté, S. Renéde Cotret, A. Juneau-Fecteau, S. Dufour-Beauséjour, M. È. Delage, D. LeBoeuf, J. Chang, B. J. Ramshaw, D. A. Bonn, W. N. Hardy, R. Liang, S. Adachi, N. E. Hussey, B. Vignolle, C. Proust, M. Sutherland, S. Krämer, J. H. Park, D. Graf, N. Doiron-Leyraud, and L. Taillefer, Nat Commun **5** (2014).

# Supplemental Material for
# "Accessing the Full Superconducting Dome in Pristine YBa$_2$Cu$_3$O$_{6+x}$ Under Pressure"

### PRESSURE-TO-HOLE DOPING TRANSLATION

Denoting $\xi(P) \equiv (1 - V(P)/V_0)$ as the relative decrease in cell volume, we start with

$$n_{\rm h}(P) = n_{\rm h,0} + \frac{dn_{\rm h}(P)}{d\xi(P)}\xi(P), \tag{3}$$

where $n_{\rm h,0} = n_{\rm h}(P=0)$. Using the first-order Murnaghan equation of state [48], $V(P)/V_0 = (1 + B_0' P/B_0)^{-1/B_0'}$, where $B_0$ is the bulk modulus $-V(dP/dV)$, $B_0'$ is the pressure derivative of $B_0$, and $V_0$ is the cell volume at zero pressure, gives the following expressions, as presented in the main text,

$$\frac{dn_{\rm h}(P)}{d\xi(P)} = n_{\rm h}'(P) B_0 \left(1 + \frac{B_0'}{B_0}P\right)^{1+1/B_0'} \tag{4}$$

$$\xi(P) = 1 - \left(1 + \frac{B_0'}{B_0}P\right)^{-1/B_0'}, \tag{5}$$

where $n_{\rm h}'(P) = \frac{dn_{\rm h}(P)}{dP}$. Substitution of expressions 4 and 5 into equation 3 gives a first-order differential equation of $n_{\rm h}(P)$ with constant parameters $B_0$, $B_0'$, and $n_{\rm h,0}$, which can be solved without further approximation with the specified boundary condition. The initial doping at zero applied pressure $n_{\rm h}(0) = 0.19$ holes/Cu was found by locating $T_{\rm c}$ on the La$_{2-x}$Sr$_x$CuO$_4$ (LSCO) superconducting parabola, $1 - T_{\rm c}/T_{\rm c,max} = 82.6(n_{\rm h} - 0.16)^2$ with $T_{\rm c,max} = 94.3$ K for YBa$_2$Cu$_3$O$_{6+x}$ (YBCO) single crystals [17]. Values considered for $B_0$ and $B_0'$ were taken from previous pressure studies using X-ray analysis, on samples with oxygen content close to that of our single crystals $x = 0.98$; as such, we assumed a range of $B_0 = 1306 \pm 11$ kbar and $B_0' = 4.8 \pm 0.2$ to $B_0 = 1500$ and $B_0' = 4$ kbar, obtained for $x = 0.95$ [49] and $x = 1$ [50], respectively. We separately set these values as parameters in expressions 4 and 5 to estimate the contribution to uncertainty in doping as a function of pressure, the horizontal error bars in figure 3 in the main text, due to errors in the exact values of $B_0$ and $B_0'$ for the oxygen content of $x = 0.98$. A median set of values was determined in this range to use for translation of the square points in figure 3.

Based on the phase diagram of YBCO up to the slightly overdoped region of $n_{\rm h} \approx 0.22$ by calcium substitution [44], we assume that the progression of $T_{\rm c}$ from the optimally doped region initially follows the LSCO parabola. Considering these constraints as the boundary condition, we solve equation 3 numerically and find that regardless of which single point we choose to constrain in accordance with this condition, the remaining data consistently appear to closely follow the LSCO parabola. Horizontal error bars in figure 3 for higher dopings reflect considerations of the possible doping points at which any possible deviation from the parabola could start to occur, as well as the uncertainty in values of $B_0$ and $B_0'$ for our exact doping, though the effect of the latter is small in comparison. Calculation using the numerical result obtained for $n_{\rm h}(P)$ shows $dn_{\rm h}(P)/d\xi(P)$ to be approximately constant along $P$, where denoting $\Delta[dn_{\rm h}(P)/d\xi(P)] \equiv dn_{\rm h}(P)/d\xi(P) - dn_{\rm h}(P=0)/d\xi(P)$, we find that $\Delta[dn_{\rm h}(P)/d\xi(P)]/[dn_{\rm h}(P=0)/d\xi(P)] \sim 0.01\%$ for $P = 0$ kbar to 300 kbar, the full pressure range considered; observing the behavior of the two factors in equation 4, it is apparent that the decrease in $n_{\rm h}'(P)$ from the numerical solution for $n_{\rm h}(P)$ effectively cancels the increase in $(1 + B_0'P/B_0)^{1+1/B_0'}$. This confirms the starting assumption of $n_{\rm h}(P)$ being linearly dependent on the relative decrease in cell volume.